\documentclass{IEEEtran}
\usepackage{graphicx}
\usepackage{amsmath}
\usepackage{cite}
\usepackage{booktabs}
\usepackage{multirow}
\usepackage{caption}
\usepackage{subcaption}
\usepackage{xcolor,colortbl}
\usepackage{svg}
\usepackage{amsfonts}
\usepackage{amssymb}

\title{Towards Precision Healthcare: Robust Fusion of Time Series and Image Data}
\author{Ali Rasekh$^{1}$, Reza Heidari$^{2}$, Amir Hosein Haji Mohammad Rezaie$^{2*}$, Parsa Sharifi Sedeh$^{2*}$, Zahra Ahmadi$^{1}$, Prasenjit Mitra$^{1}$, Wolfgang Nejdl$^{1}$\\
    $^{1}$\{ali.rasekh, ahmadi, mitra, nejdl\}@l3s.de\\
    $^{2}$\{r4heidari, haji80as, sharifyparsa1381\}@gmail.com\\
    $^{*}$Amir Hosein Haji Mohammad rezaie and Parsa Sharifi Sedeh contributed equally to this work
    }

\definecolor{Gray}{gray}{0.85}
\newcolumntype{a}{>{\columncolor{Gray}}l}

\begin{document}
\maketitle

\begin{abstract}
With the increasing availability of diverse data types, particularly images and time series data from medical experiments, there is a growing demand for techniques designed to combine various modalities of data effectively. Our motivation comes from the important areas of predicting mortality and phenotyping where using different modalities of data could significantly improve our ability to predict. To tackle this challenge, we introduce a new method that uses two separate encoders, one for each type of data, allowing the model to understand complex patterns in both visual and time-based information. Apart from the technical challenges, our goal is to make the predictive model more robust in noisy conditions and perform better than current methods. We also deal with imbalanced datasets and use an uncertainty loss function, yielding improved results while simultaneously providing a principled means of modeling uncertainty. Additionally, we include attention mechanisms to fuse different modalities, allowing the model to focus on what's important for each task. We tested our approach using the comprehensive multimodal MIMIC dataset, combining MIMIC-IV and MIMIC-CXR datasets. Our experiments show that our method is effective in improving multimodal deep learning for clinical applications. The code will be made available online.
\end{abstract}

\section{Introduction}
\label{sec:introduction}

Artificial intelligence (AI) has become increasingly essential in medical fields, transforming healthcare by offering advanced capabilities in predicting mortality, identifying diseases, and conducting various diagnostic tasks. With the rise of deep learning techniques, AI has shown outstanding effectiveness and accuracy, especially in medical applications. Multimodal learning, a recent advancement, uses various data sources like electronic health records and medical images to strengthen predictive modeling and diagnostic abilities.

The integration of AI in medical practice brings several benefits. Firstly, it allows healthcare professionals to use large amounts of data to make quick and accurate decisions. For example, in predicting mortality, AI algorithms can analyze patient data such as vital signs, lab results, and medical images to identify signs of deteriorating health and take timely action. Secondly, AI supports personalized medicine by identifying patient groups with distinct characteristics, helping tailor treatment strategies. This personalized approach improves patient outcomes and reduces the risk of adverse reactions to treatments. Additionally, AI-powered diagnostic tools are highly sensitive and specific, aiding in early disease detection. Overall, integrating AI into medical practice holds great promise for improving patient care, simplifying processes, and ultimately saving lives.

However, seamlessly integrating different types of data, like medical images and time series data, brings significant challenges in realizing AI's full potential in healthcare. Combining these diverse data types requires innovative approaches to address the complexities and variations within clinical datasets. For example, when diagnosing complex conditions like sepsis, integrating data from multiple sources such as physiological measurements and imaging studies is crucial for accurate diagnosis and timely intervention. Overcoming these integration challenges is essential for unlocking the transformative power of AI in healthcare and maximizing its impact on patient outcomes.

This paper addresses these challenges in critical healthcare tasks such as predicting mortality and phenotyping. Our main goal is to design a robust and flexible multimodal framework capable of handling the complexities of clinical datasets effectively. Ultimately, we aim to contribute to improved patient outcomes and more informed healthcare decision-making.

The rich diversity of clinical data, characterized by its multimodal nature, requires innovative approaches to extract meaningful insights. Our research focuses on achieving the following key objectives:

\begin{enumerate}
\item \textbf{Enhanced Modality Fusion via Attention Mechanism:} We introduced an attention mechanism enabling dynamic allocation of attention across modalities, enhancing model flexibility and improving predictive accuracy. This underscores the importance of modality fusion in multimodal architectures.
\item \textbf{Uncertainty-Aware Multi-Task Learning with Uncertainty Loss:} Employing an uncertainty loss function for multi-task phenotype classification, our approach prioritizes simpler and more certain tasks, enhancing overall performance by adapting to complex and uncertain ones. 
\item \textbf{Robustness in Noisy Environments:} Develop methods to ensure robust performance even in noisy settings commonly encountered in real-world hospital scenarios, where data may exhibit variability and imperfections.
\end{enumerate}

Our research yields strong results, demonstrating the practical usefulness and resilience of our multimodal framework under challenging conditions, including noisy settings and data. We chose the MIMIC dataset \cite{johnson2019mimic,johnson2023mimic} for its diverse modalities and comprehensive nature, encompassing electronic health records, time series data, and chest X-ray images. The objectives of our study, including predicting mortality, identifying diseases, and labeling radiology images, align closely with the rich annotations and labels available within the MIMIC dataset. This consolidation facilitates a comprehensive understanding of patient health, allowing our model to learn correlations between temporal health records and visual representations. The resulting multimodal dataset is carefully preprocessed, aligning timestamps and standardizing imaging data, ensuring a coherent fusion of modalities for robust model training and evaluation.

Our thorough exploration of multimodal deep neural networks for clinical applications has revealed impactful high-level findings. The specialized encoders designed for images and time series data can successfully capture patterns within each modality, significantly enhancing the model's discriminative power. Additionally, the integration of attention mechanisms for modality fusion empowers our model to allocate attention dynamically based on task and modality relevance. This not only improves adaptability but also enhances interpretability across predicting mortality and phenotyping labels such as chronic kidney diseases, other liver diseases, and complications of surgical procedures or medical care. Another critical aspect, particularly relevant in multi-label classification, is the consideration of uncertainty and how to model it effectively. We have shown that the uncertainty loss function not only improves performance but also provides a principled means of modeling, specifically in identifying diseases. Our approach showcases strong results, demonstrating the practical usefulness and resilience of our multimodal framework, even under challenging conditions such as noisy settings.

In the following sections, we first review existing work in multimodal learning and machine learning methods in healthcare data (Section \ref{sec:related}). Then, we provide a comprehensive overview of our dataset, detailing its composition, preprocessing steps, and rationale for integrating the MIMIC-IV and MIMIC-CXR datasets, following the approach outlined in the MedFuse paper\cite{hayat2022medfuse}. We detail the methodologies guiding our multimodal model training in (Section \ref{sec:method}). Subsequent sections cover experimental outcomes and our findings, highlighting the robustness and superior performance of our approach compared to state-of-the-art methods (Section \ref{sec:exper}), and conclude the paper by outlining future directions in multimodal deep learning for healthcare (Section \ref{sec:conc}).

\section{Related Work}\label{sec:related}

The field of multimodal learning has seen significant attention for its potential to extract richer representations from heterogeneous data. Our focus in this paper lies at the intersection of diverse endeavors seeking to capitalize on synergies among different data modalities. We explore two key domains in the following:  Section \ref{sec:related-ML} explores multimodal machine learning, emphasizing approaches integrating disparate data modalities for enhanced model performance. In Section \ref{sec:realted-health}, we focus on machine learning applications in healthcare, reviewing studies that have led computational methods in improving patient outcomes.

\subsection{Multimodal Machine Learning}\label{sec:related-ML}

Researchers have been exploring various methodologies to leverage multimodal data effectively in machine learning tasks. Rahim et al. \cite{RAHIM2023363}, for instance, highlight the importance of integrating longitudinal MRI images with clinical data for disease progression prediction. Their study underscores the potential of combining imaging data with other clinical variables to achieve more accurate predictions.

In a similar work, Niu et al. \cite{niu2023deep} propose a joint modeling strategy that integrates time series and clinical data to enhance mortality prediction. Their approach aims to provide a more holistic understanding of the underlying patterns influencing patient outcomes. However, the image modality is not among the modalities in their dataset.

Soenksen et al. \cite{soenksen2022integrated} take a step further by developing a platform capable of generating embeddings for various modalities, including images, text, and tabular data. Addressing the need for attention mechanisms in multimodal learning, Qiao et al. \cite{qiao2019mnn} introduce a multimodal attention mechanism that effectively incorporates textual information into learning scenarios. This approach demonstrates the benefits of attending to different modalities dynamically, leading to enhanced model performance.

In the realm of anomaly detection, Hsieh et al. \cite{hsieh2023mdf} propose a modular method inspired by Mask-RCNN for modeling both images and medical records in lung nodule detection. By combining information from multiple sources, their approach shows promise in improving the accuracy of anomaly detection tasks.

Addressing the issue of missing modalities in multimodal datasets, Lee et al. \cite{lee2023multimodal} propose a method that incorporates modality-missing aware prompts to handle missing data during training and testing. Zhang et al. \cite{zhang2022m3care} tackle the problem of missing data by leveraging auxiliary information from similar patients to impute task-related information for missing modalities in the latent space. This strategy offers a viable solution for dealing with missing data, especially when task-related information is absent.

Wang et al. \cite{wang2020multimodal} introduce a knowledge distillation-based framework that leverages modality-specific information through a teacher-student approach. By distilling knowledge from a teacher model to a student model, their approach leads to improved generalization across tasks, demonstrating the effectiveness of knowledge transfer in multimodal learning.

Ma et al. \cite{ma2022multimodal} explore the behavior of Transformers with modal-incomplete data and propose a method for searching optimal fusion strategies via multi-task optimization. Their approach demonstrates dataset-dependent optimal fusion strategies and improved model robustness, albeit with acknowledged limitations for safety-critical systems.

In the domain of medical imaging, Nie et al. \cite{nie20163d} propose a deep learning model for Alzheimer's Disease (AD) diagnosis that integrates MRI and PET data through separate CNNs. Enhanced with a novel correlation calculation method, their approach outperforms other models, showcasing promise for superior diagnostic efficiency in AD.

Nie et al. \cite{nie2019multi} introduce a two-stage learning method for predicting overall survival (OS) time in high-grade glioma patients. By leveraging a multi-channel 3D CNN to integrate various MRI modalities followed by an SVM stage, they achieve high accuracy in predicting OS time. In addition, Srinivas and Sasibhushana Rao \cite{srinivas2020segmentation} present a comprehensive two-stage learning method focused on predicting overall survival (OS) time in high-grade glioma patients. This approach integrates a multi-channel 3D CNN, combining contrast-enhanced T1 MRI, diffusion tensor imaging (DTI), and resting-state functional MRI (rs-fMRI). Following this, the method assimilates features with select demographic and tumor-related features into a support vector machine (SVM). The results underscore the potential of deep learning in neuro-oncological applications for individualized treatment planning.

Muduli et al. \cite{muduli2022automated} present a deep convolutional neural network (CNN) model for automated breast cancer classification using mammograms and ultrasound images. Their streamlined architecture demonstrates effective feature extraction, yielding superior classification performance compared to existing models, highlighting the potential of automated systems for accurate breast cancer detection and early diagnosis.

Introducing a multimodal deep belief network (DBN), Liang et al. \cite{liang2014integrative} showcase its effectiveness in integrative analysis of multi-platform cancer data. Their study emphasizes the practical impact of this approach in cancer pathogenesis studies, offering valuable insights for personalized treatment strategies.

Sun et al. \cite{sun2018multimodal} propose a multimodal deep neural network by integrating multi-dimensional data (MDNNMD) for predicting breast cancer prognosis. Their study achieves superior performance compared to existing approaches, emphasizing the potential of deep learning and multi-dimensional data integration for improving breast cancer prognosis prediction.

Developing a multimodal deep learning model, Joo et al. \cite{joo2021multimodal} focus on predicting the pathologic complete response (pCR) to neoadjuvant chemotherapy (NAC) in breast cancer patients. Integrating clinical information and high-dimensional MR images, their model significantly outperforms models using only clinical information or MR images, showcasing the potential of deep learning in combining diverse information sources for improved pCR prediction.

In their research, Khan et al. \cite{khan2023multi} propose a multimodal deep neural network for multi-class malignant liver diagnosis. Integrating portal venous computed tomography (CT) scans and pathology data, they employ transfer learning to address insufficient and imbalanced datasets. Their study demonstrates superior performance compared to existing liver diagnostic studies, offering a significant contribution to malignant liver diagnosis.

\subsection{Machine Learning in Healthcare}\label{sec:realted-health}
Despite promising advancements in applying machine learning to healthcare, significant limitations persist, including the need for extensive labeled data, interpretability challenges, and potential biases. This section examines these issues and current research efforts aimed at addressing them to ensure ethical and practical implications are carefully considered.

Combining structured clinical data with unstructured clinical narratives from electronic health records (EHR), Zeng et al. \cite{zeng2019identifying} tackle the prediction of distant recurrences in breast cancer. Similarly, Harerimana et al. \cite{harerimana2019deep} explores the potential of deep learning in leveraging Electronic Health Records for clinical tasks, offering intuitive explanations and blueprints for health informatics professionals to apply deep learning algorithms effectively in medical settings.

Addressing the needs of type 1 diabetes patients, Smith et al. \cite{jeon2020predicting} focus on predicting blood glucose concentration, employing sophisticated data imputation techniques and various feature engineering methods.

Navigating missing longitudinal clinical data in ICU laboratory test results, Smith et al. \cite{daberdaku2020combined} propose a hybrid approach, combining linear interpolation with weighted K-Nearest Neighbors (KNN) methods. Their method outperforms existing techniques in most analytes, showcasing robustness in clinical data imputation.

In the realm of predictive modeling for inpatient mortality among Medicare patients, researchers \cite{zikos2021cross} enhance Healthcare Cost and Utilization Project (HCUP) tools with acuity and diagnosis presence information. This augmentation not only refines predictive models but also offers insights into factors influencing inpatient mortality rates.

Zhu et al. \cite{zhu2020dilated} introduce DeepFall, a pioneering fall detection framework leveraging deep spatio-temporal convolutional autoencoders. Their method outperforms traditional approaches on publicly available datasets, promising advancements in privacy-preserving fall detection mechanisms.

Tekiroğlu and Erkan \cite{turkmen2023bioberturk} contribute to the field by developing Turkish biomedical language models, specifically introducing the BioBERTurk family and a labeled dataset for head CT radiology report classification. Their work enhances classification performance in clinical contexts, catering to linguistic diversity within medical data analysis.

Recent studies have emphasized the vital role of chest X-rays (CXR) in predicting mortality and ventilatory support requirements for COVID-19 patients. Balbi et al. \cite{balbi2021chest} particularly highlight CXR as a complementary tool alongside clinical and demographic data for early prognosis and risk assessment. This underscores the importance of integrating imaging modalities with traditional diagnostic approaches to enhance patient care in critical conditions like COVID-19.

Machine learning techniques, particularly those leveraging Big Data Analytics, have shown promise in disease prediction and prognosis. Mir and Dhage \cite{mir2018diabetes} explored the prediction of diabetes using machine learning algorithms, evaluating their performance on the Pima Indians Diabetes Database. Their study highlights the potential of machine learning in enabling early disease diagnosis and facilitating informed decision-making in healthcare, thereby enhancing patient outcomes.

Maity and Das \cite{maity2017machine} delved into the applications of machine learning in healthcare, with a specific focus on diagnosis and prognosis. Through case studies on Alzheimer's disease diagnosis and breast cancer severity classification, Maity demonstrates how machine learning algorithms can efficiently analyze complex healthcare data, enabling timely intervention and personalized treatment strategies.

In the domain of oncology, Rahane \cite{rahane2018lung} explored the use of image processing and machine learning techniques for lung cancer detection. By employing SVM and image processing on CT scan images and blood samples, Rahane's study underscores the importance of early detection in improving treatment outcomes for lung cancer patients. This highlights the critical role of advanced imaging modalities and computational techniques in enhancing cancer care and patient survival rates.

Moving to hepatology, Gogi and Vijayalakshmi \cite{gogi2018prognosis} explored the use of machine learning algorithms for liver disease prognosis. By analyzing Liver Function Tests (LFT) parameters using Decision Tree, Linear Discriminant, SVM Fine Gaussian, and Logistic Regression algorithms, Gogi demonstrates the potential of machine learning in enhancing prognosis methods for liver diseases.

In the domain of disease risk prediction, Shuai et al. \cite{niu2021label} proposed a method that integrates attention mechanisms with Clinical-BERT to improve interpretability in disease risk prediction using textual inputs. Their approach focuses on joint embedding of words and labels, effectively utilizing medical notes to enhance prediction accuracy while maintaining interpretability.

Jacenkow et al. \cite{jacenkow2022indication} explored the influence of textual information, specifically the indication field in radiology reports, on image classification tasks. By fine-tuning a transformer network pre-trained on text data for multimodal classification, they achieved improved performance in identifying image features relevant to clinical indications.

Bezirganyan et al. \cite{bezirganyan2023m2} introduced M2-Mixer, a novel architecture for multimodal classification tasks. Their approach, based on MLP-Mixer, incorporates a multi-head loss function to address modality predominance issues. By simplifying the architecture and introducing multi-head loss, they achieved superior performance compared to existing baseline models on benchmark datasets.

Guo et al. \cite{guo2021predicting} focused on predicting mortality among patients with liver cirrhosis using electronic health record (EHR) data and machine learning techniques. By comparing deep learning models with traditional Models for End Stage Liver disease (MELD) scores, they showcased the superior performance of deep learning in mortality prediction.

Suvon et al. \cite{suvon2022multimodal} tackled the challenging task of predicting mortality in patients with Pulmonary Arterial Hypertension (PAH) using a multimodal learning approach. By integrating features from numerical imaging, echo report categorical features, and echo report text features extracted using BERT, they developed a comprehensive framework for mortality prediction. Their experiments showed significant improvements in prediction accuracy.

The paper MedFuse\cite{hayat2022medfuse} is the most related work to ours. This research introduces a novel LSTM-based fusion module designed to integrate uni-modal and multimodal input, addressing challenges in multimodal fusion where data collection is asynchronous. MedFuse demonstrates significant performance improvements in in-hospital mortality prediction and phenotype classification tasks by using clinical time series data and chest X-ray images from the MIMIC-IV and MIMIC-CXR datasets. Unlike other fusion approaches, MedFuse treats multimodal representations as a sequence of uni-modal representations, performing even with partially paired data. However, the selected architecture is not the best-performing and robust solution. Their model also does not consider the uncertainty of the different tasks in multi-task phenotyping classification. Our proposed approach, including our architecture, attention mechanism, preprocessing methods, and the incorporation of uncertainty modeling through a loss function, enabled us to outperform their method in both noisy and noise-free environments.

\section{Multimodal Data Preparation}\label{sec:data}

This section describes the process of preparing our multimodal dataset. First, we will discuss the details of the MIMIC dataset (Section~\ref{subsec:mimic_dataset}), including its overall structure and relevant information for our analysis. Next, we will explore the specifics of the MIMIC-CXR subset (Section~\ref{subsec:mimic-cxr}), which focuses on chest X-ray images within the MIMIC dataset. Finally, we will review the MIMIC-IV dataset (Section~\ref{subsec:mimic-iv}), which provides additional information relevant to our study. At the end (Section~\ref{subsec:multimodal data generation}), we'll discuss how the data from MIMIC-IV and MIMIC-CXR are combined to generate the multimodal dataset.

\subsection{MIMIC dataset}
\label{subsec:mimic_dataset}
The Medical Information Mart for Intensive Care (MIMIC) is a large, freely available database containing healthcare data. MIMIC-IV (v2.2) \cite{johnson2023mimic}, the latest version released in January 2023, incorporates data from patients admitted between the years 2008 and 2019. It improves upon numerous aspects of its predecessors by adopting a modular data organization approach, highlighting data provenance. This section provides a high-level overview of the MIMIC-IV dataset, highlighting its structure and relevant information for our analysis. MIMIC-IV's rich data allows researchers to explore a wide range of healthcare topics, including patient demographics, diagnoses, procedures, medications, laboratory measurements, vital signs, and even information from the online medical record system (e.g., height and weight). Importantly, the data is deidentified using a strict protocol to protect patient privacy while still enabling valuable medical research.

\subsection{MIMIC-CXR}
\label{subsec:mimic-cxr}

MIMIC-CXR \cite{johnson2019mimic} is a valuable subset of MIMIC specifically focusing on chest X-ray images. This publicly available resource provides a rich dataset for researchers in medical image analysis. It contains 377,110 de-identified chest radiographs, including both frontal and lateral views captured during patients' hospital admissions. In this research paper, the multimodal dataset utilized combines MIMIC-CXR with MIMIC-IV, integrating chest X-ray images from MIMIC-CXR as the second modality. This fusion allows for a comprehensive dataset including both time series data and image modalities.

\subsection{MIMIC-IV}
\label{subsec:mimic-iv}

The MIMIC-IV database represents a significant advancement in medical data resources, building upon the success of MIMIC-III. Incorporating contemporary data from 2008 to 2019, MIMIC-IV is sourced from two in-hospital database systems, a custom hospital-wide Electronic Health Record (EHR) and an ICU-specific clinical information system. The latest version of this dataset includes the information of 299,712 patients, 431,231 admissions, and 73,181 ICU stays. This dataset is used to run experiments on the proposed model for two tasks; namely in-hospital mortality prediction, and phenotyping task. The latter includes 25 binary labels for predicting a range of diseases which can be categorized into groups of acute, mixed, and chronic diseases. For instance, the \textit{chronic kidney diseases} label in the chronic group is considered for patients who have long-term damage to the kidneys, often progressive and irreversible, leading to impaired kidney function. As an acute disease, \textit{complications of surgical procedures or medical care} are Adverse events or complications arising from surgical or medical treatments and \textit{other liver diseases} label determines the various liver disorders not classified under specific categories, including conditions like fatty liver disease and hepatitis. MIMIC-IV serves as a valuable resource for driving advancements in clinical informatics, epidemiology, and machine learning to improve patient care and outcomes. In our study, we have used this dataset to get the time series modality for our research.

\subsection{Generating Multimodal Dataset}\label{subsec:multimodal data generation}
As previously mentioned, we employ the MIMIC-IV dataset to assess and benchmark our models. To generate this dataset, we adopt the data extraction and preprocessing approach proposed in the MedFuse\cite{hayat2022medfuse}. In the original dataset, each patient report may contain zero or more chest X-ray images taken during the patient's hospital stay, in addition to time series data. Our data generation process involves selecting the last captured image for each patient report and combining it with the associated time series data to create a sample.

We utilize consistent dataset settings for reporting our results. Employing the patient identifier from the clinical time series data, we randomly partition the dataset into 70\% for training, 10\% for validation, and 20\% for the test set. In our notation, we denote the clinical time series data as EHR and the chest X-ray images as CXR. The dataset is categorized into (EHR+CXR)PARTIAL, containing paired and partially paired samples (i.e., samples with missing chest X-rays), and (EHR + CXR)PAIRED, containing data samples where both modalities are present.

For instance, the (EHR + CXR)PARTIAL training set for patient phenotyping comprises 7,756 samples, with chest X-rays included among 42,628 samples. Chest X-ray images are extracted from MIMIC-CXR and split based on a random patient split. Images from the training set are then transferred to either the validation or test set if associated with patients in the validation or test splits of the clinical time series data. This procedure results in 325,188 images in the training set, 15,282 images in the validation set, and 36,625 images in the test set.

\begin{figure*} 
    \centering
  \subfloat[]{\includegraphics[width=0.25\textwidth]{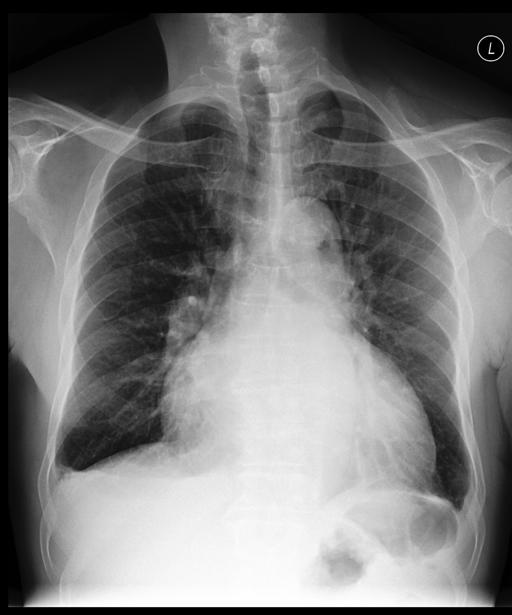}}
  \subfloat[]{\includegraphics[width=0.25\textwidth]{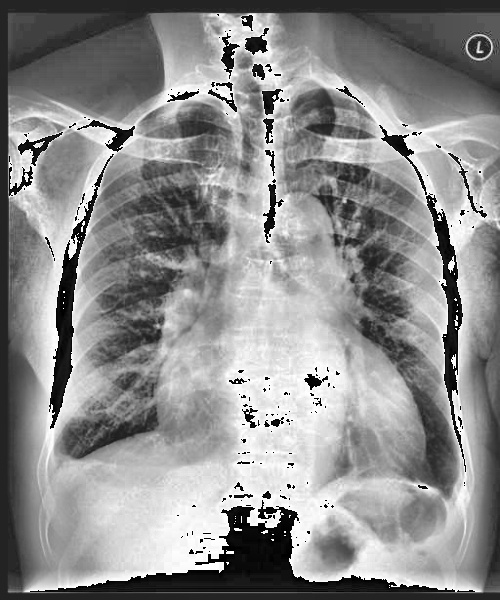}}

    \caption{The effect of CLAHE augmentation: (a) original chest X-ray image, (b) CLAHE augmented image. CLAHE significantly improves the visibility of inner body parts, showcasing intricate details such as the kidney on the right side of the image. Additionally, it enhances the depiction of bone density, providing clearer insights. Such refined details play a crucial role in mortality prediction and phenotype classification. }\label{fig:CLAHE_AUG}
\end{figure*}

\subsection{Preprocessing}
We use the MedFuse\cite{hayat2022medfuse} data extraction and preprocessing procedure along with another preprocessing method for images known as CLAHE\cite{clahe}. For chest X-ray images, a consistent set of transformations is applied during both pre-training and fine-tuning across all experiments and tasks. Specifically, each image is resized to 256 × 256 pixels, undergoes a random horizontal flip, and experiences various random affine transformations, including rotation, scaling, shearing, and translation. Subsequently, a random crop is applied to achieve an image size of 224 × 224 pixels. During the validation and testing phases, image resizing to 256 × 256 and a center crop to 224 × 224 pixels are performed.

To ensure fair comparisons and showcase the efficacy of multimodal learning, we utilize a consistent set of 17 clinical variables. 
Among these, five are categorical: capillary refill rate, glasgow coma scale eye opening, glasgow coma scale motor response, glasgow coma scale verbal response, and glasgow coma scale total. The remaining 12 are continuous variables: diastolic blood pressure, fraction of inspired oxygen, glucose, heart rate, height, mean blood pressure, oxygen saturation, respiratory rate, systolic blood pressure, temperature, weight, and pH. The input for all tasks is regularly sampled every two hours, with discretization and standardization of clinical variables following established protocols, as detailed in prior work.

After data preprocessing and one-hot encoding of categorical features, we obtain a vector representation of size 76 at each time step in the clinical time series data. For a given instance, the representation is denoted as $x_{ehr} \in \mathbb{R}^{t\times76}$, where the value of $t$ is dependent on the specific instance and task.

In addition, we explored a data preprocessing method known as CLAHE contrast enhancement. Developed as an extension of traditional histogram equalization, CLAHE provides a dynamic and localized approach to contrast enhancement. The method involves dividing an image into small, non-overlapping tiles and independently applying histogram equalization to each tile. This adaptive approach ensures that contrast enhancement is tailored to the unique characteristics of local regions, preventing the over-amplification of noise in homogeneous areas. We applied the CLAHE method to all images to enhance image quality, thereby facilitating the extraction of more information by the model.
In Figure.\ref{fig:CLAHE_AUG} the impact of CLAHE on the image is visible. One of the primary benefits of CLAHE in chest X-ray imaging is its ability to enhance the visibility of lung parenchyma. In addition to lung pathology, CLAHE enhances the visualization of thoracic skeletal structures, including the ribs, vertebrae, and mediastinum. This improved depiction of bony anatomy is invaluable for detecting fractures, degenerative changes, and mediastinal masses, thereby assisting in the diagnosis of conditions ranging from traumatic injuries to neoplastic processes.

\begin{figure*}[h]%
\centering
\includegraphics[width=0.9\textwidth]{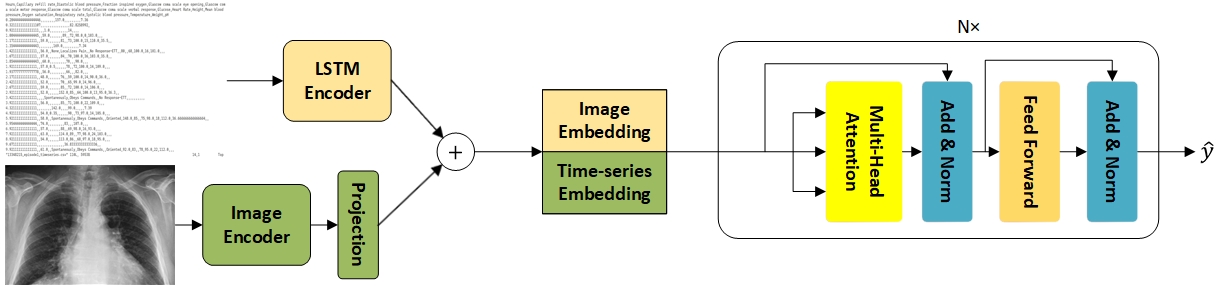}
\caption{Model architecture consisting of modality-specific encoders and a multilayer transformer encoder as our multimodal fusion network. }\label{fig:model_arch}
\end{figure*}

\section{Proposed Framework}\label{sec:method}

In the following sections, we will describe our proposed framework for multimodal healthcare analysis and prediction. We will explore the architecture of our model including the modality-specific encoders and the attention mechanism used for modality fusion (Section \ref{sec:method-arch}). Then we will go into more detail about our attention-based model in Section \ref{subsec:Attention-Based Multimodal Fusion}. Finally, we will explain the motives for using the uncertainty loss function and its advantages (Section \ref{subsec:Uncertainty Loss}).

\subsection{Model Architecture}\label{sec:method-arch}
Our proposed model shown in Figure \ref{fig:model_arch} consists of two major parts: modality-specific encoders and a multimodal Transformer encoder \cite{NIPS2017_3f5ee243} as our modality fusion network. We use an image encoder (e.g., a ResNet-34 model \cite{7780459}) to extract features from our image modality and an LSTM network \cite{HochSchm97} to extract latent feature representations from our time series modality. We then utilize a projection layer to project the image embeddings to the time series embedding dimension. Finally, we concatenate these feature representations and feed them to a Transformer encoder to predict the labels.

In the first part, we pre-train our encoders to independently extract meaningful representations from each of our modalities. An LSTM architecture works best for extracting feature embeddings from our time series data due to the quantity of available data and its consecutive nature. We use the Adam optimizer \cite{Kingma2014AdamAM} to optimize our Binary Cross-Entropy losses and pre-train our modality-specific encoders.

We use a ResNet-34 model as the backbone for our image encoder and we set the output dimension of its classifier layer to be equal to the number of labels in our specific task. For our time series backbone, we use an LSTM network with $N = 2$ layers stacked on top of each other with a hidden dimension of $d = 256$ and a dropout layer with a dropout probability of $p = 0.3$. We also utilize a linear layer as the final classifier for our LSTM network.

In the second part, we dismiss the classifiers for our encoders and use latent feature embeddings $f^{’}_{cxr}$ and $f_{ehr}$. We feed $f^{’}_{cxr}$ to a fully connected projection layer to get $f_{cxr}$ that has the same dimensionality as $f_{ehr}$. We then concatenate $f_{ehr}$ and $f_{cxr}$ to create the sequence $f_{fused}$ that consists of our unimodal feature embeddings:
\begin{equation}
    f_{fused} = [f_\text{ehr}, projection(f^{'}_\text{cxr})].
\end{equation}

We use a Transformer Encoder without positional embeddings with a linear layer on top of that as our modality fusion network to resolve the issue of modality bias in our baselines. In models that use an LSTM network for fusion a major problem is the order of modality embeddings in the input sequence. Therefore, the sequence order may create a bias towards the modalities that come first in the input sequence, and changing it may vary results significantly when in reality the modalities do not possess a specific ordering. By using an attention-based network for fusion our model learns the importance of every modality in each of our tasks and thus performs better than state-of-the-art LSTM or MLP architectures. Finally, we optimize the multi-task uncertainty loss \eqref{eq_uncertainty} introduced in Kendall et al. \cite{kendall2018multi} with an Adam optimizer to fine-tune our network.

\subsection{Attention-Based Multimodal Fusion}\label{subsec:Attention-Based Multimodal Fusion}
We feed our concatenated unimodal feature embedding $f_{fused}$ to a Transformer Encoder network with $L = 2$ encoder layers stacked on top of one another, each of them having $h = 8$ heads and a feedforward dimension of $c = 1024$. We then feed the output of this network to a linear classifier to get the final predictions $\hat{y}$. Finally, we optimize the following multi-task uncertainty loss function with an Adam optimizer to fine-tune our model:
\begin{equation}
    L(\hat{y}, y) = UncertaintlyLoss(\hat{y}, y).
\end{equation}

Our attention-based model can learn the importance of each modality for each task simultaneously and find meaningful relations between the latent features of different modalities. This allows our model to deeply integrate different modalities to better understand our task and achieve more precise results.

\subsection{Uncertainty Loss}
\label{subsec:Uncertainty Loss}
A key issue with our baseline models is their dependency on the relative weights of each task's loss. For instance, in phenotype classification we have 25 different labels (each label corresponds to a separate task) and usually, the same weight is given to each of their losses, or the weights are manually selected. Giving the same weights to the losses of different tasks could negatively impact our model’s performance due to the separate nature of each task. Manually choosing the relative weights is also a time-consuming ordeal that should be done for every single classification problem separately and requires vast resources. 

\begin{figure*}[h]%
\centering
\includegraphics[width=0.9\textwidth]{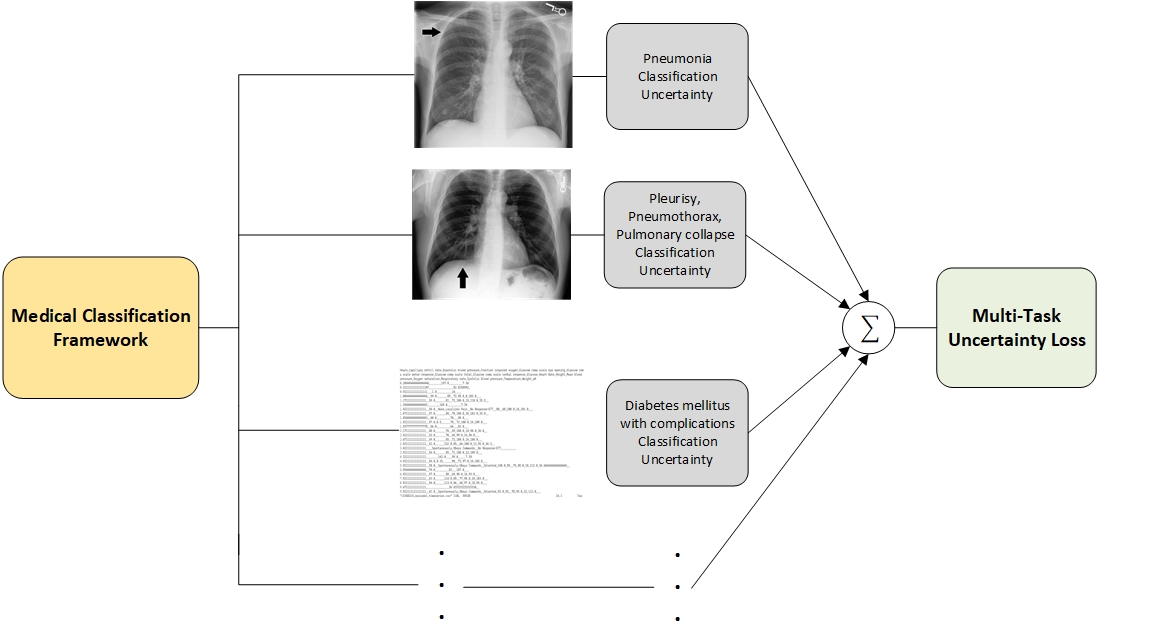}
\caption{We combine and weigh multiple losses according to the uncertainty of each task to compute the multi-task uncertainty loss.}\label{fig:multi_task_uncertainty}
\end{figure*}

The multi-task uncertainty loss shown in Figure \ref{fig:multi_task_uncertainty} largely resolves this issue by weighing multiple losses and considering the homoscedastic uncertainty of each task. This method learns the weighing parameter $\sigma$ for each of the losses during the training process. The multi-task uncertainty loss function for classification is the following:
\begin{equation}
    UncertaintlyLoss(\hat{y}, y) = \sum\limits_{i=1}^N\frac{1}{\sigma_i^2}L(\hat{y}, y) + log \sigma_i^2,
\label{eq_uncertainty}\end{equation}
where $L(\hat{y}, y)$ is the Binary Cross-Entropy loss and $N$ is the number of tasks.

By using this loss to fine-tune our model and learn its parameters while simultaneously learning the value of $\sigma$ for each task, we were able to boost the performance of our model in phenotype classification and achieve better results.

\section{Experiments and Results}\label{sec:exper}
In this section, we explain our experiments. Starting from the complex architecture described in the method section, we explain how we put it into practice with a detailed demonstration of our experimental setup and baseline models. We experimented with our method on two important medical tasks: in-hospital mortality prediction and phenotypes classification. After analyzing these tasks, we break down the different parts of our approach to see how each contributes. Additionally, we investigate uncertainties and test the model's robustness.

\subsection{Experimental Setup}\label{subsec2}
In this study, we establish our experimental framework using the MIMIC-IV and the  MIMIC-CXR datasets. Our choice of MIMIC is based on its extensive scale, comprehensive documentation and standardized formatting. Our experiments focus on two key objectives:

\begin{enumerate}
\item Predicting the binary in-hospital mortality label after the first 48 hours of a patient's ICU stay.
\item Classify a set of 25 phenotype labels for the patients during their ICU stays.

\end{enumerate}

We train our proposed network separately for each task and evaluate the results. We use a batch size of $16$ for our data loader. For each task, we first pre-train our ResNet-34 encoder with images from the MIMIC-CXR dataset, and pre-train the LSTM encoder network with time series data from the MIMIC-IV dataset. 

For the phenotyping task, we use learning rates of $5 \times 10^{-4}$ and $1 \times 10^{-4}$ for pre-training the image and time series encoders, respectively and for the in-hospital mortality task, we use learning rates of $5 \times 10^{-4}$ and $3 \times 10^{-5}$ for pre-training the image and time series encoders. Then we jointly fine-tune the fusion module with the encoders, using a learning rate of $7 \times 10^{-5}$ for the phenotyping task and $1 \times 10^{-4}$ for the in-hospital mortality task.

\subsection{Baselines}
\textbf{MedFuse} \cite{hayat2022medfuse}  is a distinctive approach involving the utilization of a fusion module based on an LSTM architecture. This module effectively combines information from both image and time series data. Fine-tuning of the MedFuse model follows a two-step pre-training process. Initially, the image encoder is pre-trained on 14 radiology labels, focusing on the detection of a specific disease in unpaired chest X-rays. Simultaneously, the LSTM is pre-trained using unpaired time series data for in-hospital mortality prediction or phenotype classification. It's worth noting that, despite the use of unpaired data during pre-training, MedFuse requires labels for a distinct task in the image modality. For experimental evaluation, the publicly available MedFuse model is applied to the multimodal MIMIC dataset. The dataset is partitioned into similar splits as those employed in our work and other baseline models, ensuring a fair comparison.

\textbf{Contrastive-based} \cite{simClr}
Contrastive learning is a machine learning paradigm that aims to teach a model the differences and similarities between different data modalities. The fundamental idea behind contrastive learning is to embed similar samples closer to each other in a latent space while pushing dissimilar samples apart. In establishing this baseline, we have adopted an approach similar to that in \cite{inter_intra_modality}, wherein we construct a model with 2 headers, using a ViT-Base image encoder \cite{dosovitskiy2021an} and an LSTM time series encoder as the backbone of the model. One header is designated for inter-modality optimization, while the other optimizes the intra-modality loss. For every pair of image and time series data, we apply random augmentations and try to maximize the similarity of the data and the augmented version. We also employ the other header to align the representations of the image and time series more closely.

\textbf{Diffusion-based classifier}
In the domain of classifiers utilizing diffusion mechanisms, our literature review revealed an absence of pre-existing multimodal diffusion-based classifiers. We extended the recent unimodal diffusion-based classifier CARD \cite{han2022card} to a multimodal version. The original architecture employs an encoder(ResNet-34) to convert image data samples into prior vectors. Subsequently, the diffusion backward process, facilitated by a denoising deep neural network, aims to denoise these prior vectors, refining them into more accurate feature vectors for classification. The final denoised prior vectors are then used for classification. To create a multimodal diffusion-based classifier, we replaced the CARD encoder with an encoder concatenating the separate embeddings of image and time series data.

\subsection{In-Hospital Mortality Prediction}\label{subsec2}

For the critical task of in-hospital mortality prediction, we evaluate the performance of our proposed model against established baselines, including MedFuse, CARD, and Contrastive Learning. Table \ref{tab:mortality} presents a comparative analysis of macro average F1-score, binary F1-score, AUROC, and AUPRC metrics. Our model consistently outperforms state-of-the-art approaches, demonstrating its efficacy in predicting mortality across diverse modalities.

Upon reviewing the performance of the models, it’s clear that the CARD model falls short, demonstrating the lowest performance among all our baselines. The contrastive model, which employs the ViT architecture, surpasses MedFuse across all metrics, with the exception of AUROC. Our multimodal attention-based model outperforms MedFuse and CARD on all metrics and surpasses the contrastive model on macro average F1-score and AUROC. By training our proposed model with the CLAHE augmentation on images, we achieve superior results in all metrics excluding the macro average F1-score, where the filter slightly decreases the performance of our attention-based model.

For this experiment we use time series data and chest X-ray images, to predict the in-hospital mortality of the patients according to the first 48 hours of ICU stay in a binary classification task. There are 18845 samples in our training set, 2138 samples in our validation set and 5243 samples in our test set. All samples include time series data, but 4885, 540 and 1373 samples have both modalities in the training, validation and test sets, respectively. The rest of the samples in each set have missing chest X-ray images.

\begin{table*}[t] %
\caption{In-hospital mortality prediction performance.}
\centering %
\begin{tabular}{@{}lllll@{}} %
\hline
Model & Macro Average F1-score  & Binary F1-score & AUROC & AUPRC \\
\hline
MedFuse    & 0.677   & 0.412  & 0.857 & 0.507  \\
Contrastive + ViT & 0.690   & 0.441  & 0.852 & 0.520  \\
CARD & 0.660 & 0.400 & 0.690 & 0.341  \\
Attention &   \textbf{0.691} & 0.438  & 0.857 & 0.514  \\
Attention + CLAHE & 0.685 & \textbf{0.453} & \textbf{0.858} & \textbf{0.524} \\
\hline
\end{tabular}

\label{tab:mortality}
\end{table*}

\begin{table*}[t]
\caption{Phenotype classification performance.}%
\centering
\begin{tabular}{@{}lllll@{}}
\hline
Model & Macro Average F1-score  & Binary F1-score & AUROC & AUPRC \\
\hline
MedFuse    & 0.589   & 0.282  & 0.763 & 0.422  \\
CARD    & 0.585   & 0.344 & 0.600 & 0.310\\
Attention + CLAHE & 0.611   & 0.327  & \textbf{0.770} & 0.431  \\
Attention + Uncertainty + CLAHE & \textbf{0.614} &  \textbf{0.362}  & 0.759 & \textbf{0.466}  \\
\hline
\end{tabular}
\label{tab:phenotyping}%
\end{table*}

\begin{table*}[t]
\caption{Ablation study results.}%
\centering
\begin{tabular}{@{}lllll@{}}
\hline
Model & Macro Average F1-score  & Binary F1-score & AUROC & AUPRC \\
\hline
Time series only  & 0.581   & 0.270   & 0.759  & 0.421  \\
Image only &   0.535 & 0.200  & 0.670 & 0.358  \\
Multimodal LSTM Fusion  & 0.589  & 0.282   & 0.763 & 0.422  \\
Multimodal Attention & 0.604   & 0.314  & 0.765 & 0.424  \\
Attention + Uncertainty Loss & 0.604   & 0.311  & 0.767 & 0.427  \\
Attention + CLAHE & 0.611   & 0.327  & \textbf{0.770} & 0.431  \\
Attention + Uncertainty + CLAHE & \textbf{0.614} &  \textbf{0.362}  & 0.759 & \textbf{0.466}  \\
\hline
\end{tabular}
\label{tab:ablations}
\end{table*}

\begin {table*}[t]
\centering
\caption{Robustness experiment results on models trained on noisy data.}%
\begin{tabular}{@{}llllll@{}}
\hline
Model & Percentage of Noise & AUROC & AUPRC & Macro Average F1-score & Binary F1-score\\
\hline
Attention & \%10  & \textbf{0.768} & \textbf{0.426} & \textbf{0.601} & \textbf{0.306}  \\
MedFuse    & \%10   & 0.765  & 0.422 & 0.585 & 0.274 \\
\hline
Attention & \%20  & \textbf{0.763} & \textbf{0.423}  & \textbf{0.600} & \textbf{0.308}\\
MedFuse    & \%20   & 0.761  & 0.420 & 0.590 & 0.281\\
\hline
Attention & \%30  & 0.751 & \textbf{0.403} & \textbf{0.610} & \textbf{0.322}\\
MedFuse    & \%30   & \textbf{0.757}  & 0.410 & 0.591 & 0.288\\
\hline
Attention & \%40  & \textbf{0.763} & \textbf{0.421} & \textbf{0.602} & \textbf{0.310}\\
MedFuse    & \%40   & 0.757  & 0.414 & 0.594 & 0.295\\
\hline
Attention & \%50  & \textbf{0.761} & \textbf{0.417} & \textbf{0.594} & \textbf{0.291}\\
MedFuse    & \%50   & 0.757  & 0.410 & 0.583 & 0.270\\
\hline
Attention & \%60  & 0.755 & 0.412 & \textbf{0.602} & \textbf{0.312}\\
MedFuse    & \%60   & 0.755  & 0.412 & 0.580 & 0.271\\
\hline
\end{tabular}
\label{tab:robustness}
\end{table*}

\begin{table*}[t]
\centering
\caption{robustness experiment results on models trained on noise-free data.}%
\begin{tabular}{@{}llllll@{}}
\hline
Model & Percentage of Noise & AUROC & AUPRC & Macro Average F1-score & Binary F1-score \\
\hline

Attention & \%10  & \textbf{0.761} & \textbf{0.418} & \textbf{0.590} & \textbf{0.299} \\
MedFuse & \%10   & 0.757  & 0.412 & 0.581 &  0.284 \\
\hline
Attention & \%20  & 0.737 & 0.387 & 0.581 & \textbf{0.279}  \\
MedFuse & \%20   & \textbf{0.743}  & \textbf{0.395} & \textbf{0.559} & 0.239  \\
\hline
Attention & \%30  & \textbf{0.757} & \textbf{0.411} & \textbf{0.581} & \textbf{0.268}  \\
MedFuse & \%30   & 0.753  & 0.404 & 0.557 &  0.220 \\
\hline
Attention & \%40  & \textbf{0.743} & 0.396 & \textbf{0.570} & \textbf{0.246}  \\
MedFuse & \%40   & 0.724  & 0.369  & 0.557 & 0.202 \\
\hline
Attention & \%50  & \textbf{0.732} & \textbf{0.379}  & \textbf{0.561} & \textbf{0.245} \\
MedFuse & \%50   & 0.717  & 0.363 & 0.546 & 0.204  \\
\hline
Attention & \%60  & \textbf{0.689} & \textbf{0.334}  & \textbf{0.540} & \textbf{0.195} \\
MedFuse & \%60   & 0.674  & 0.315  & 0.510 & 0.153 \\
\hline
\end{tabular}
\label{tab:robustness_not_noisy}
\end{table*}

\begin{table*}
\caption{Task-wise uncertainty impacts.}
\centering
\begin{tabular}{@{}lcc|cc@{}}
\toprule
 & \multicolumn{2}{l}{Attention with Uncertainty} & \multicolumn{2}{l}{Attention without Uncertainty} \\
 \hline
Phenotype & AUROC & AUPRC & AUROC & AUPRC\\
\midrule
Acute and unspecified renal failure & 0.793 & \textbf{0.592} & 0.793 & 0.590  \\
Acute cerebrovascular disease   & 0.908 & \textbf{0.468} &  0.908 & 0.462  \\
Acute myocardial infarction  & 0.760 & 0.216  & 0.762 & 0.221   \\
Cardiac dysrhythmias    & 0.681 & \textbf{0.484} &  0.681 & 0.482 \\
Chronic kidney disease  & \textbf{0.746} & \textbf{0.439} & 0.736 & 0.429   \\
Chronic obstructive pulmonary disease and bronchiectasis  & \textbf{0.704} & \textbf{0.292}  & 0.701 & 0.289   \\
Complications of surgical procedures or medical care   & \textbf{0.731} & \textbf{0.406}  & 0.730 & 0.405   \\
Conduction disorders    & \textbf{0.718} & 0.246 &  0.717 & 0.248   \\
Congestive heart failure; nonhypertensive  & \textbf{0.765} & \textbf{0.522}  & 0.760 & 0.516   \\
Coronary atherosclerosis and other heart disease  & \textbf{0.761} & \textbf{0.591} & 0.760 & 0.588   \\
Diabetes mellitus with complications   & \textbf{0.900} & \textbf{0.595} & 0.897 & 0.592 \\
Diabetes mellitus without complication  & \textbf{0.789} & \textbf{0.413} & 0.787 & 0.406 \\
Disorders of lipid metabolism & 0.704 & \textbf{0.619} & 0.706 & 0.617 \\
Essential hypertension & \textbf{0.668} & \textbf{0.579} & 0.660 & 0.570 \\
Fluid and electrolyte disorders & 0.759 & 0.644 & 0.760 & 0.644 \\
Gastrointestinal hemorrhage  & 0.772 & 0.213 & 0.772 & 0.215 \\
Hypertension with complications and secondary hypertension & \textbf{0.738} & \textbf{0.432} & 0.728 & 0.421 \\
Other liver diseases & \textbf{0.741} & 0.297 & 0.737 & 0.307 \\
Other lower respiratory disease & 0.655 & 0.165 & 0.657 & 0.169 \\
Other upper respiratory disease  & \textbf{0.752} & \textbf{0.261} & 0.747 & 0.249 \\
Pleurisy; pneumothorax; pulmonary collapse & 0.708 & 0.155 & 0.714 & 0.158 \\
Pneumonia & 0.\textbf{813} & \textbf{0.384} & 0.811 & 0.382 \\
Respiratory failure; insufficiency; arrest (adult) & \textbf{0.872} & 0.565 & 0.871 & 0.565 \\
Septicemia (except in labor) & \textbf{0.846} & \textbf{0.522} & 0.842 & 0.512 \\
Shock & \textbf{0.890} & \textbf{0.569} & 0.888 & 0.552 \\
\hline
Average & \textbf{0.767} & \textbf{0.427} & 0.765 & 0.424 \\
\hline
\end{tabular}
\label{tab:uncertainty}
\end{table*}

\begin{figure*}[h]
  \centering
  \includegraphics[width=0.8\textwidth]{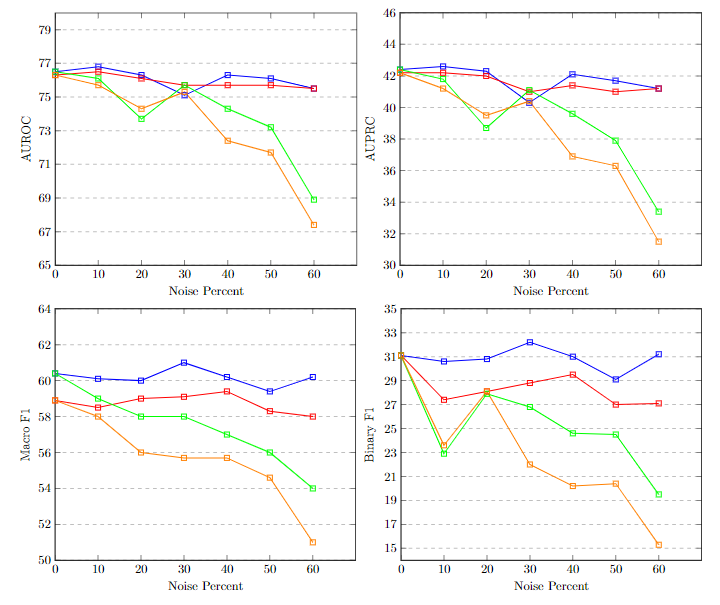}
  \caption{Performance comparison of models trained on noisy or noise-free datasets, and evaluated on noisy datasets. The plot employs different colors to represent specific configurations: 
  blue indicates the attention-based fusion model trained on noisy data; red shows the MedFuse model trained on noisy data; green denotes the attention-based fusion trained on noise-free data; and orange represents the MedFuse model trained on noise-free data. As noise levels increase, a general decline in performance is observed for all models across various metrics. Notably, the use of attention mechanisms appears to mitigate performance degradation, showcasing enhanced robustness against noise.}
  \label{fig:robustness}
\end{figure*}

\subsection{Phenotype Classification}\label{subsec2}

In the domain of phenotyping, our model is evaluated against baselines, including MedFuse, CARD, and Contrastive learning. Table \ref{tab:phenotyping} illustrates the model's proficiency in addressing the complex task of multi-label phenotyping. Metrics such as macro average F1-score, binary F1-score, AUROC, and AUPRC are employed to evaluate the model's accuracy in capturing diverse phenotypic characteristics. The results showcase the model's ability to handle the intricacies of multi-label phenotyping, outperforming the baselines.

When we compare the models, we find that the MedFuse model performs better than the CARD model in all metrics, with the exception of the binary F1-score. Our attention-based model trained on images augmented by the CLAHE filter surpasses CARD and MedFuse in macro average F1-score and AUROC but it has a lower binary F1-score compared to CARD. When we optimize our model with the uncertainty loss it outperforms all of our baselines in all metrics except for AUROC, where it appears that combining uncertainty loss with the CLAHE augmentation slightly decreases the performance of our model. It’s important to note that all the best results across various metrics were achieved by our models.

The objective of this experiment is to predict 25 different conditions given to patients during the length of their ICU stay. We utilize time series data and chest X-ray images for this multi-label classification task. 
There are 42628 samples in our training set, 4802 samples in our validation set and 11914 samples in our test set. All samples include time series data, but 7756, 882 and 2166 samples have both modalities in the training, validation and test sets. The rest of the samples in each set have missing chest X-ray images.

\subsection{Ablation Study}
The ablation study examines the model's architecture, exploring distinctive configurations to understand their impact. These configurations include variations such as using time series only, images only and fusing image and time series data through LSTM or attention mechanisms. The results are shown in Table \ref{tab:ablations}, providing insights into the contributions of the model parts.

In our primary observation, the LSTM-fused multimodal model demonstrates superior performance in phenotype classification compared to uni-modal models across all metrics. This outcome underscores the effectiveness of the multimodal approach in fusing the information derived from both modalities for classification purposes. Subsequently, the attention-based fusion model surpasses the LSTM-fused model, emphasizing the contribution of the transformer layers in enhancing the multimodal model's classification performance.

Incorporating uncertainty loss into the model yields improved performance, particularly evident in AUPRC and AUROC metrics, without significantly impacting the macro F1-score. A minor decrease (approximately 0.003) is observed in the binary F1-score. In the end, we analyzed the impact of the CLAHE filter on CXR images. Results indicate that using CLAHE improves the performance of our attention-based model across all metrics. Also, adding this filter increases our attention-based uncertainty model's performance across all metrics except for AUROC.

In summary, the comprehensive analysis presented in this table highlights the positive impact of the modules and methods employed, collectively contributing to improved model performance, particularly evident in terms of AUPRC and AUROC metrics.

\subsection{Task-wise Uncertainties}
To understand the importance of utilizing homoscedastic uncertainty in our multi-label phenotyping task, we have compared the performance of our attention-based model fine-tuned using the uncertainty loss with the same model fine-tuned using the Cross-Entropy loss. 

As you can see in Table \ref{tab:uncertainty}, The uncertainty loss not only enhances the average performance of our proposed framework but more importantly, this loss function increases the individual performance on most labels in our multi-label setup. This loss function allows our model to focus more on tasks that are easier to predict and have less uncertainty without causing a significant decrease in performance in the less uncertain and more complicated tasks, thus achieving a higher average performance. Although the overall improvements might not be considered drastic, the constant improvement in different tasks, especially those that had high performances prior to using the uncertainty loss, further supports the claim that the loss function makes our model more flexible and allows it to focus on more certain tasks, and indicates the importance of using this loss function in our multi-label setup.

\subsection{Robustness}
In order to explore the robustness of our attention-based model, we compared the performance of our model against MedFuse in noisy configurations. To do so, we prepared a noisy version of the multimodal MIMIC dataset. Table \ref{tab:robustness} presents the results across various noise levels, ranging from 10\% to 60\%, on both the training and testing sets. In this experiment, we subject all samples to varying levels of noise. For instance, in the case of images, we introduce noise by perturbing 10\% of the pixels within the data. Similarly, for the time-series data, we apply noise to 10\% of the time steps in each sample. The introduced noise is Gaussian and its mean and standard deviation are estimated by measuring these parameters in 1000 random samples of the data. These parameters are calculated individually for each feature in the time-series data and all pixels in the images.

Two different modes are considered for the evaluation process. In one, models are trained and tested on noisy datasets. In the other, testing is conducted on noisy datasets without any prior training on noisy datasets. The latter scenario is very common in real-world applications, where unexpected noise is often encountered in the data. The results for each setting are presented in Table \ref{tab:robustness} and Table \ref{tab:robustness_not_noisy}, while the corresponding Figure \ref{fig:robustness} mirrors the tabulated results. It is evident that as noise levels increase, the model accuracy decreases. However, the attention-based fusion results in superior overall performance compared to the MedFuse model.\\
Unlike the MedFuse model, which exhibits significant performance degradation in the presence of noise, our attention-based model demonstrates high levels of robustness. For instance, even at 60\% noise levels, the decrease in performance is minimal, with results showcasing as little as a 2\% reduction in performance. These results underscore the efficacy of our attention mechanism in mitigating the effects of noise, ensuring consistent and reliable performance across diverse environmental conditions.

\section{Conclusion}\label{sec:conc}
Our research introduces an innovative approach to multimodal deep neural networks, specifically designed for integrating heterogeneous data modalities such as images and time series data in mortality prediction and phenotyping label assignments. By employing dedicated encoders for each modality, our model effectively captures nuanced patterns inherent in both visual and temporal information, thus enhancing predictive capabilities. Our experiments conducted under noisy settings demonstrate the robustness of our model, surpassing state-of-the-art methods and showcasing its efficacy in handling real-world challenges with noisy clinical data. Additionally, our innovative use of an uncertainty loss function addresses the complexity of multi-label classification, contributing to improved model performance. Furthermore, the integration of attention mechanisms for modality fusion enhances adaptability by dynamically allocating attention based on task relevance. In summary, our research advances robust multimodal deep learning for clinical applications, offering a flexible and robust framework capable of addressing challenges in real-world clinical data, with promising outcomes for clinical decision support systems.

Looking ahead, several promising directions for future research emerge from the outcomes of this study. First and foremost, enhancing the interpretability of the multimodal deep neural network remains a critical area of investigation. Developing methodologies that shed light on the decision-making process of the model will be essential for building trust in clinical applications. This may involve exploring novel visualization techniques or model-agnostic interpretability tools to dissect how the network integrates information from diverse modalities, providing clinicians with valuable insights into its decision rationale.

Our research sets the stage for exploring the integration of additional data modalities into our multimodal deep neural network framework. While our current focus has been on combining images and time series data for mortality prediction and phenotyping, there's immense potential in extending our approach to include other modalities such as textual data or genomic information. By incorporating these additional modalities, we can investigate how our model performs across a broader spectrum of clinical data and further enhance its predictive capabilities. This exploration opens up avenues for understanding how different types of data interact and contribute to the overall predictive power of our model, helping to have more comprehensive and robust clinical decision support systems.

\clearpage

\bibliographystyle{unsrt}
\bibliography{sn-bibliography}

\begin{thebibliography}{10}

\bibitem{johnson2019mimic}
Alistair~EW Johnson, Tom~J Pollard, Nathaniel~R Greenbaum, Matthew~P Lungren, Chih-ying Deng, Yifan Peng, Zhiyong Lu, Roger~G Mark, Seth~J Berkowitz, and Steven Horng.
\newblock Mimic-cxr-jpg, a large publicly available database of labeled chest radiographs.
\newblock {\em arXiv preprint arXiv:1901.07042}, 2019.

\bibitem{johnson2023mimic}
Alistair~EW Johnson, Lucas Bulgarelli, Lu~Shen, Alvin Gayles, Ayad Shammout, Steven Horng, Tom~J Pollard, Sicheng Hao, Benjamin Moody, Brian Gow, et~al.
\newblock Mimic-iv, a freely accessible electronic health record dataset.
\newblock {\em Scientific data}, 10(1):1, 2023.

\bibitem{hayat2022medfuse}
Nasir Hayat, Krzysztof~J Geras, and Farah~E Shamout.
\newblock Medfuse: Multi-modal fusion with clinical time-series data and chest x-ray images.
\newblock In {\em Machine Learning for Healthcare Conference}, pages 479--503. PMLR, 2022.

\bibitem{RAHIM2023363}
Nasir Rahim, Shaker El-Sappagh, Sajid Ali, Khan Muhammad, Javier {Del Ser}, and Tamer Abuhmed.
\newblock Prediction of alzheimer's progression based on multimodal deep-learning-based fusion and visual explainability of time-series data.
\newblock {\em Information Fusion}, 92:363--388, 2023.

\bibitem{niu2023deep}
Ke~Niu, Ke~Zhang, Xueping Peng, Yijie Pan, and Naian Xiao.
\newblock Deep multi-modal intermediate fusion of clinical record and time series data in mortality prediction.
\newblock {\em Frontiers in Molecular Biosciences}, 10:1136071, 2023.

\bibitem{soenksen2022integrated}
Luis~R Soenksen, Yu~Ma, Cynthia Zeng, Leonard Boussioux, Kimberly Villalobos~Carballo, Liangyuan Na, Holly~M Wiberg, Michael~L Li, Ignacio Fuentes, and Dimitris Bertsimas.
\newblock Integrated multimodal artificial intelligence framework for healthcare applications.
\newblock {\em NPJ digital medicine}, 5(1):149, 2022.

\bibitem{qiao2019mnn}
Zhi Qiao, Xian Wu, Shen Ge, and Wei Fan.
\newblock Mnn: multimodal attentional neural networks for diagnosis prediction.
\newblock {\em Extraction}, 1(2019):A1, 2019.

\bibitem{hsieh2023mdf}
Chihcheng Hsieh, Isabel~Blanco Nobre, Sandra~Costa Sousa, Chun Ouyang, Margot Brereton, Jacinto~C Nascimento, Joaquim Jorge, and Catarina Moreira.
\newblock Mdf-net: Multimodal dual-fusion network for abnormality detection using cxr images and clinical data.
\newblock {\em arXiv preprint arXiv:2302.13390}, 2023.

\bibitem{lee2023multimodal}
Yi-Lun Lee, Yi-Hsuan Tsai, Wei-Chen Chiu, and Chen-Yu Lee.
\newblock Multimodal prompting with missing modalities for visual recognition.
\newblock In {\em Proceedings of the IEEE/CVF Conference on Computer Vision and Pattern Recognition}, pages 14943--14952, 2023.

\bibitem{zhang2022m3care}
Chaohe Zhang, Xu~Chu, Liantao Ma, Yinghao Zhu, Yasha Wang, Jiangtao Wang, and Junfeng Zhao.
\newblock M3care: Learning with missing modalities in multimodal healthcare data.
\newblock In {\em Proceedings of the 28th ACM SIGKDD Conference on Knowledge Discovery and Data Mining}, pages 2418--2428, 2022.

\bibitem{wang2020multimodal}
Qi~Wang, Liang Zhan, Paul Thompson, and Jiayu Zhou.
\newblock Multimodal learning with incomplete modalities by knowledge distillation.
\newblock In {\em Proceedings of the 26th ACM SIGKDD International Conference on Knowledge Discovery \& Data Mining}, pages 1828--1838, 2020.

\bibitem{ma2022multimodal}
Mengmeng Ma, Jian Ren, Long Zhao, Davide Testuggine, and Xi~Peng.
\newblock Are multimodal transformers robust to missing modality?
\newblock In {\em Proceedings of the IEEE/CVF Conference on Computer Vision and Pattern Recognition}, pages 18177--18186, 2022.

\bibitem{nie20163d}
Dong Nie, Han Zhang, Ehsan Adeli, Luyan Liu, and Dinggang Shen.
\newblock 3d deep learning for multi-modal imaging-guided survival time prediction of brain tumor patients.
\newblock In {\em Medical Image Computing and Computer-Assisted Intervention--MICCAI 2016: 19th International Conference, Athens, Greece, October 17-21, 2016, Proceedings, Part II 19}, pages 212--220. Springer, 2016.

\bibitem{nie2019multi}
Dong Nie, Junfeng Lu, Han Zhang, Ehsan Adeli, Jun Wang, Zhengda Yu, LuYan Liu, Qian Wang, Jinsong Wu, and Dinggang Shen.
\newblock Multi-channel 3d deep feature learning for survival time prediction of brain tumor patients using multi-modal neuroimages.
\newblock {\em Scientific reports}, 9(1):1103, 2019.

\bibitem{srinivas2020segmentation}
B~Srinivas and Gottapu Sasibhushana~Rao.
\newblock Segmentation of multi-modal mri brain tumor sub-regions using deep learning.
\newblock {\em Journal of Electrical Engineering \& Technology}, 15(4):1899--1909, 2020.

\bibitem{muduli2022automated}
Debendra Muduli, Ratnakar Dash, and Banshidhar Majhi.
\newblock Automated diagnosis of breast cancer using multi-modal datasets: A deep convolution neural network based approach.
\newblock {\em Biomedical Signal Processing and Control}, 71:102825, 2022.

\bibitem{liang2014integrative}
Muxuan Liang, Zhizhong Li, Ting Chen, and Jianyang Zeng.
\newblock Integrative data analysis of multi-platform cancer data with a multimodal deep learning approach.
\newblock {\em IEEE/ACM transactions on computational biology and bioinformatics}, 12(4):928--937, 2014.

\bibitem{sun2018multimodal}
Dongdong Sun, Minghui Wang, and Ao~Li.
\newblock A multimodal deep neural network for human breast cancer prognosis prediction by integrating multi-dimensional data.
\newblock {\em IEEE/ACM transactions on computational biology and bioinformatics}, 16(3):841--850, 2018.

\bibitem{joo2021multimodal}
Sunghoon Joo, Eun~Sook Ko, Soonhwan Kwon, Eunjoo Jeon, Hyungsik Jung, Ji-Yeon Kim, Myung~Jin Chung, and Young-Hyuck Im.
\newblock Multimodal deep learning models for the prediction of pathologic response to neoadjuvant chemotherapy in breast cancer.
\newblock {\em Scientific reports}, 11(1):18800, 2021.

\bibitem{khan2023multi}
Rayyan~Azam Khan, Minghan Fu, Brent Burbridge, Yigang Luo, and Fang-Xiang Wu.
\newblock A multi-modal deep neural network for multi-class liver cancer diagnosis.
\newblock {\em Neural Networks}, 165:553--561, 2023.

\bibitem{zeng2019identifying}
Zexian Zeng, Liang Yao, Ankita Roy, Xiaoyu Li, Sasa Espino, Susan~E Clare, Seema~A Khan, and Yuan Luo.
\newblock Identifying breast cancer distant recurrences from electronic health records using machine learning.
\newblock {\em Journal of healthcare informatics research}, 3:283--299, 2019.

\bibitem{harerimana2019deep}
Gaspard Harerimana, Jong~Wook Kim, Hoon Yoo, and Beakcheol Jang.
\newblock Deep learning for electronic health records analytics.
\newblock {\em IEEE Access}, 7:101245--101259, 2019.

\bibitem{jeon2020predicting}
Jouhyun Jeon, Peter~J Leimbigler, Gaurav Baruah, Michael~H Li, Yan Fossat, and Alfred~J Whitehead.
\newblock Predicting glycaemia in type 1 diabetes patients: experiments in feature engineering and data imputation.
\newblock {\em Journal of healthcare informatics research}, 4(1):71--90, 2020.

\bibitem{daberdaku2020combined}
Sebastian Daberdaku, Erica Tavazzi, and Barbara Di~Camillo.
\newblock A combined interpolation and weighted k-nearest neighbours approach for the imputation of longitudinal icu laboratory data.
\newblock {\em Journal of Healthcare Informatics Research}, 4:174--188, 2020.

\bibitem{zikos2021cross}
Dimitrios Zikos, Aashara Shrestha, and Leonidas Fegaras.
\newblock A cross-sectional study to predict mortality for medicare patients based on the combined use of hcup tools.
\newblock {\em Journal of Healthcare Informatics Research}, pages 1--19, 2021.

\bibitem{zhu2020dilated}
Taiyu Zhu, Kezhi Li, Jianwei Chen, Pau Herrero, and Pantelis Georgiou.
\newblock Dilated recurrent neural networks for glucose forecasting in type 1 diabetes.
\newblock {\em Journal of Healthcare Informatics Research}, 4:308--324, 2020.

\bibitem{turkmen2023bioberturk}
Hazal T{\"u}rkmen, O{\u{g}}uz Dikenelli, Cenk Eraslan, Mehmet~Cem Call{\i}, and S{\"u}ha~S{\"u}reyya {\"O}zbek.
\newblock Bioberturk: Exploring turkish biomedical language model development strategies in low-resource setting.
\newblock {\em Journal of Healthcare Informatics Research}, pages 1--14, 2023.

\bibitem{balbi2021chest}
Maurizio Balbi, Anna Caroli, Andrea Corsi, Gianluca Milanese, Alessandra Surace, Fabiano Di~Marco, Luca Novelli, Mario Silva, Ferdinando~Luca Lorini, Andrea Duca, et~al.
\newblock Chest x-ray for predicting mortality and the need for ventilatory support in covid-19 patients presenting to the emergency department.
\newblock {\em European radiology}, 31:1999--2012, 2021.

\bibitem{mir2018diabetes}
Ayman Mir and Sudhir~N Dhage.
\newblock Diabetes disease prediction using machine learning on big data of healthcare.
\newblock In {\em 2018 fourth international conference on computing communication control and automation (ICCUBEA)}, pages 1--6. IEEE, 2018.

\bibitem{maity2017machine}
Niharika~G Maity and Sreerupa Das.
\newblock Machine learning for improved diagnosis and prognosis in healthcare.
\newblock In {\em 2017 IEEE aerospace conference}, pages 1--9. IEEE, 2017.

\bibitem{rahane2018lung}
Wasudeo Rahane, Himali Dalvi, Yamini Magar, Anjali Kalane, and Satyajeet Jondhale.
\newblock Lung cancer detection using image processing and machine learning healthcare.
\newblock In {\em 2018 International Conference on Current Trends towards Converging Technologies (ICCTCT)}, pages 1--5. IEEE, 2018.

\bibitem{gogi2018prognosis}
Vyshali~J Gogi and MN~Vijayalakshmi.
\newblock Prognosis of liver disease: Using machine learning algorithms.
\newblock In {\em 2018 International Conference on Recent Innovations in Electrical, Electronics \& Communication Engineering (ICRIEECE)}, pages 875--879. IEEE, 2018.

\bibitem{niu2021label}
Shuai Niu, Qing Yin, Yunya Song, Yike Guo, and Xian Yang.
\newblock Label dependent attention model for disease risk prediction using multimodal electronic health records.
\newblock In {\em 2021 IEEE International Conference on Data Mining (ICDM)}, pages 449--458. IEEE, 2021.

\bibitem{jacenkow2022indication}
Grzegorz Jacenk{\'o}w, Alison~Q O’Neil, and Sotirios~A Tsaftaris.
\newblock Indication as prior knowledge for multimodal disease classification in chest radiographs with transformers.
\newblock In {\em 2022 IEEE 19th International Symposium on Biomedical Imaging (ISBI)}, pages 1--5. IEEE, 2022.

\bibitem{bezirganyan2023m2}
Grigor Bezirganyan, Sana Sellami, Laure Berti-{\'E}Quille, and S{\'e}bastien Fournier.
\newblock M2-mixer: A multimodal mixer with multi-head loss for classification from multimodal data.
\newblock In {\em 2023 IEEE International Conference on Big Data (BigData)}, pages 1052--1058. IEEE, 2023.

\bibitem{guo2021predicting}
Aixia Guo, Nikhilesh~R Mazumder, Daniela~P Ladner, and Randi~E Foraker.
\newblock Predicting mortality among patients with liver cirrhosis in electronic health records with machine learning.
\newblock {\em PloS one}, 16(8):e0256428, 2021.

\bibitem{suvon2022multimodal}
Mohammod~NI Suvon, Prasun~C Tripathi, Samer Alabed, Andrew~J Swift, and Haiping Lu.
\newblock Multimodal learning for predicting mortality in patients with pulmonary arterial hypertension.
\newblock In {\em 2022 IEEE International Conference on Bioinformatics and Biomedicine (BIBM)}, pages 2704--2710. IEEE, 2022.

\bibitem{clahe}
Garima Yadav, Saurabh Maheshwari, and Anjali Agarwal.
\newblock Contrast limited adaptive histogram equalization based enhancement for real time video system.
\newblock In {\em 2014 International Conference on Advances in Computing, Communications and Informatics (ICACCI)}, pages 2392--2397, 2014.

\bibitem{NIPS2017_3f5ee243}
Ashish Vaswani, Noam Shazeer, Niki Parmar, Jakob Uszkoreit, Llion Jones, Aidan~N Gomez, \L~ukasz Kaiser, and Illia Polosukhin.
\newblock Attention is all you need.
\newblock In I.~Guyon, U.~Von Luxburg, S.~Bengio, H.~Wallach, R.~Fergus, S.~Vishwanathan, and R.~Garnett, editors, {\em Advances in Neural Information Processing Systems}, volume~30. Curran Associates, Inc., 2017.

\bibitem{7780459}
Kaiming He, Xiangyu Zhang, Shaoqing Ren, and Jian Sun.
\newblock Deep residual learning for image recognition.
\newblock In {\em 2016 IEEE Conference on Computer Vision and Pattern Recognition (CVPR)}, pages 770--778, 2016.

\bibitem{HochSchm97}
Sepp Hochreiter and Jürgen Schmidhuber.
\newblock Long short-term memory.
\newblock {\em Neural Computation}, 9(8):1735--1780, 1997.

\bibitem{Kingma2014AdamAM}
Diederik~P. Kingma and Jimmy Ba.
\newblock Adam: A method for stochastic optimization.
\newblock {\em CoRR}, abs/1412.6980, 2014.

\bibitem{kendall2018multi}
Alex Kendall, Yarin Gal, and Roberto Cipolla.
\newblock Multi-task learning using uncertainty to weigh losses for scene geometry and semantics.
\newblock In {\em Proceedings of the IEEE conference on computer vision and pattern recognition}, pages 7482--7491, 2018.

\bibitem{simClr}
Ting Chen, Simon Kornblith, Mohammad Norouzi, and Geoffrey Hinton.
\newblock A simple framework for contrastive learning of visual representations, 2020.

\bibitem{inter_intra_modality}
Xin Yuan, Zhe Lin, Jason Kuen, Jianming Zhang, Yilin Wang, Michael Maire, Ajinkya Kale, and Baldo Faieta.
\newblock Multimodal contrastive training for visual representation learning, 2021.

\bibitem{dosovitskiy2021an}
Alexey Dosovitskiy, Lucas Beyer, Alexander Kolesnikov, Dirk Weissenborn, Xiaohua Zhai, Thomas Unterthiner, Mostafa Dehghani, Matthias Minderer, Georg Heigold, Sylvain Gelly, Jakob Uszkoreit, and Neil Houlsby.
\newblock An image is worth 16x16 words: Transformers for image recognition at scale.
\newblock In {\em International Conference on Learning Representations}, 2021.

\bibitem{han2022card}
Xizewen Han, Huangjie Zheng, and Mingyuan Zhou.
\newblock Card: Classification and regression diffusion models.
\newblock {\em Advances in Neural Information Processing Systems}, 35:18100--18115, 2022.

\end{thebibliography}

\end{document}